\documentclass[runningheads]{llncs}
\usepackage{graphicx}
\usepackage{amsfonts}
\usepackage[table]{xcolor}
\usepackage{subfig}\usepackage{color}

\usepackage{float}
\usepackage{booktabs}
\usepackage{mathtools}
\usepackage{multirow}
\usepackage{tabularx}
\usepackage{dirtytalk}
\usepackage{hyperref}

\newcommand{\vect}[1]{\boldsymbol{#1}}
\newcolumntype{R}[1]{>{\raggedleft\let\newline\\\arraybackslash\hspace{0pt}}p{#1}}

\begin{document}
%
\title{Calibration window selection based on change- point detection for forecasting electricity prices\thanks{Supported by the Ministry of Education \& Science (MEiN, Poland) through Grant No. 0027/DIA/2020/49 (to WN) and the National Science Center (NCN, Poland) through Grant No. 2018/30/A/HS4/00444 (to RW). \textit{Forthcoming in}: Proceedings of the International Conference on Computational Science (ICCS) 2022, London, UK.}}
\titlerunning{Calibration window selection based on change-point detection}
%
\author{Julia Nasiadka
\and
Weronika Nitka\thanks{Corresponding author}
\and
Rafa{\l} Weron}
%
%
\institute{Department of Operations Research and Business Intelligence, \\
Wroc{\l}aw University of Science and Technology, 50-370 Wroc{\l}aw, Poland 
\email{weronika.nitka@pwr.edu.pl}}
\maketitle              
\begin{abstract}
We employ a recently proposed change-point detection algorithm, the Narrowest-Over-Threshold (NOT) method, to select subperiods of past observations that are similar to the currently recorded values. Then, contrarily to the traditional time series approach in which the most recent $\tau$ observations are taken as the calibration sample, we estimate autoregressive models only for data in these subperiods. We illustrate our approach using a challenging dataset -- day-ahead electricity prices in the German EPEX SPOT market -- and observe a significant improvement in forecasting accuracy compared to commonly used approaches, including the Autoregressive Hybrid Nearest Neighbors (ARHNN) method.
\end{abstract}

\section{Introduction}

Electricity price forecasting (EPF) is an extremely challenging task. A number of methods have been developed for this purpose,  ranging from linear regression to hybrid deep learning architectures utilizing long-short term memory and/or convolutional neural networks. While most studies focus on improving model structures, selecting input features with more predictive power or implementing more efficient algorithms \cite{Heijden2021119954,Jahangir20202369,lago_forecasting_2021}, the issue of the optimal calibration window is generally overlooked \cite{hubicka_note_2019a}.

This work is inspired by a recent article \cite{DeMarcos2020}, which utilized a relatively simple change-point detection method \cite{Zeileis2003109} to split the time series into segments with the `same' price level, and an ICCS 2021 paper \cite{nitka_forecasting_2021a}, which employed the $k$-nearest neighbors ($k$-NN) algorithm to select the calibration sample based on similarity over a subset of explanatory variables. Here, we utilize a recently proposed change-point detection algorithm -- the Narrowest-Over-Threshold (NOT) method \cite{baranowski_NOT_2019} -- to construct an automatic method for detecting subperiods exhibiting different temporal dynamics. Once  identified, those not resembling the current behavior are discarded when estimating the predictive model. In what follows, we provide empirical evidence that significant improvement in forecasting accuracy can be achieved compared to commonly used EPF approaches.

The remainder of the paper is structured as follows. In Section \ref{sec:data} we present the dataset and the transformation, which is used to standardize the data. In Section \ref{sec:NOT} we briefly describe the NOT method and introduce our approach to selecting subperiods for model calibration. Next, in Section \ref{sec:methods} we present the forecasting models and in Section \ref{sec:results} the empirical results. Finally, in Section \ref{sec:conclusions} we conclude and discuss future research directions.

\begin{figure}[tb]
\centering
\includegraphics[width=.9\textwidth]{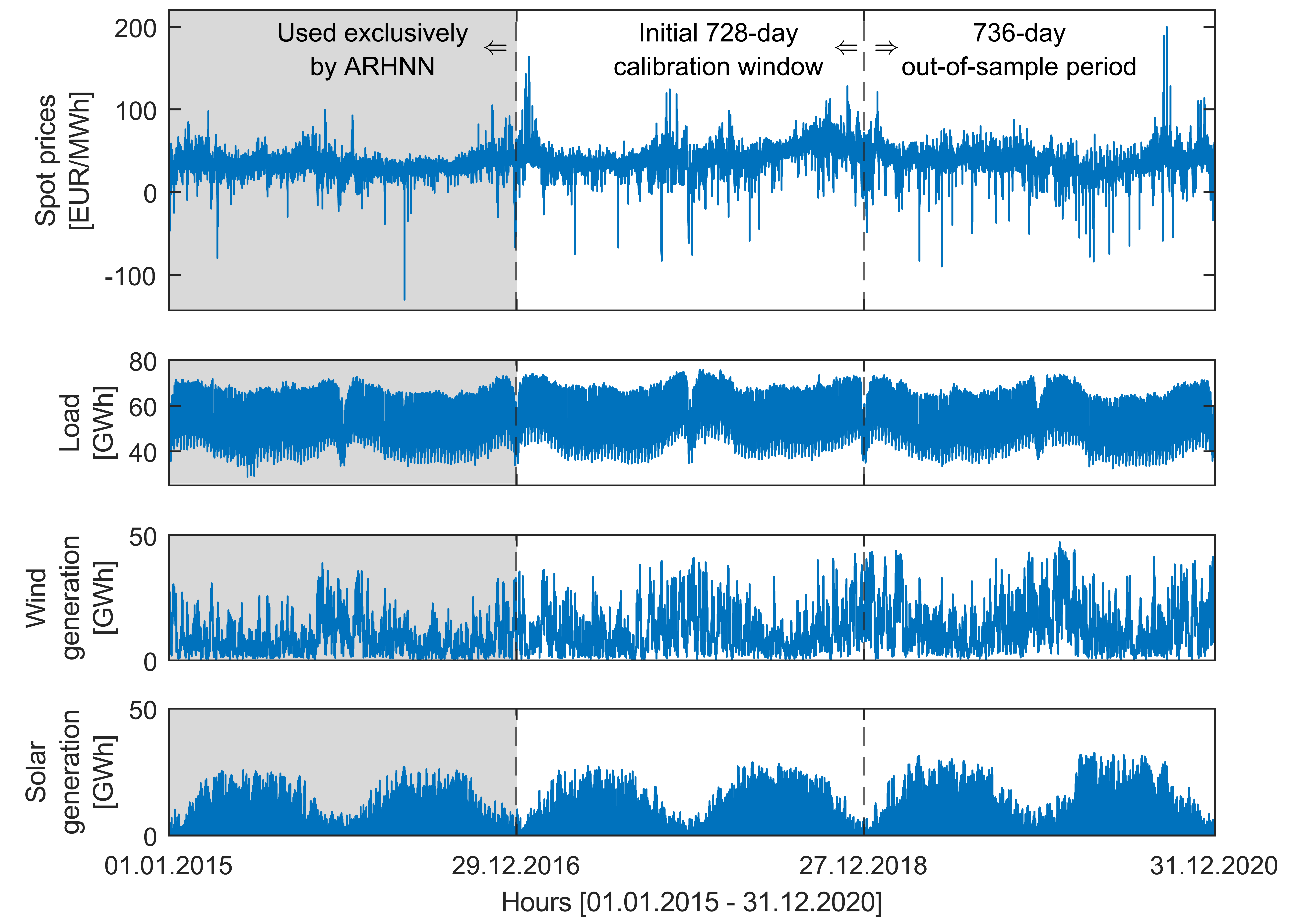}
\caption{Electricity spot prices and day-ahead load, wind and solar power generation forecasts in Germany. The last 736 days constitute the test period.}
\label{fig:data_de}
\end{figure}

\section{The data}
\label{sec:data}

For comparison purposes, we use the same dataset as in \cite{nitka_forecasting_2021a}. It spans six years (2015-2020) at hourly resolution and includes four series from the German EPEX SPOT market: electricity spot prices $P_{d,h}$ (more precisely: prices set in the day-ahead auction on day $d-1$ for the 24h of day $d$) and day-ahead load $\hat{L}_{d,h}$, wind $\hat{W}_{d,h}$ and solar power generation $\hat{S}_{d,h}$ forecasts, see Fig.\ \ref{fig:data_de}. The first two years are exclusively used for estimating the Autoregressive Hybrid Nearest Neighbors (ARHNN) method \cite{nitka_forecasting_2021a}; the remaining methods require less data for calibration. The last 736 days constitute the out-of-sample test period.

The most distinct feature of the German power market are frequent spikes and negative prices. Similarly volatile is wind energy generation, while load and solar generation are more predictable. Following \cite{uniejewski_variance_2018,ziel_dayahead_2018a}, to cope with this extreme volatility, we transform the electricity prices using the area hyperbolic sine:
$Y_{d,h} = \operatorname{asinh}{(\frac{1}{b}\{P_{d,h}-a\})}$ with $\operatorname{asinh}{(x)} = \log\{x + (x^2 + 1)^{0.5}\}$, where $a$ is the median of $P_{d,h}$ in the calibration window and $b$ is median absolute deviation. The price forecasts are then obtained by the inverse transformation. 

Note that in \cite{nitka_forecasting_2021a} a different transformation was used. 
All series, not just $P_{d,h}$, were normalized by subtracting the mean and dividing by the standard deviation in each calibration window. We denote models utilizing \textit{asinh}-transformed data with subscript H; the remaining ones use the standard normalization, as in \cite{nitka_forecasting_2021a}.

\begin{figure}[tb]
\centering
\includegraphics[width=.9\textwidth]{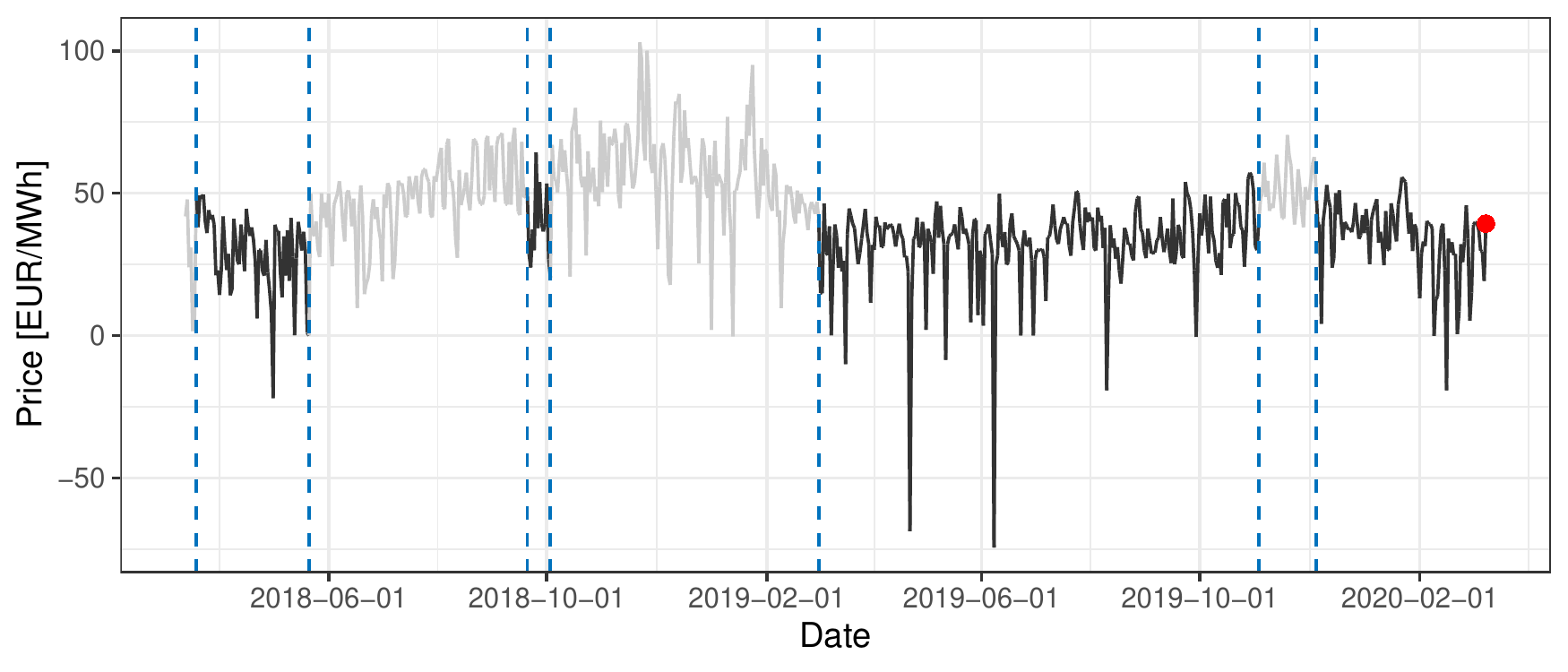}
\caption{A sample run of the algorithm introduced in Section \ref{sec:NOT}. The red dot is the target day. Vertical dashed lines indicate the located change-points separating periods with different statistical properties. The discarded prices are in gray.}
\label{fig:window_selection}
\end{figure}

\section{Calibration window selection using NOT}
\label{sec:NOT}



The Narrowest-Over-Threshold (NOT) method \cite{baranowski_NOT_2019} can detect an unknown number of change-points at unknown locations in one-dimensional time series data. The key feature is its focus on the smallest local sections of the data on which the existence of a change-point is suspected. A change-point is said to occur when the behavior of the series changes significantly \cite{DeMarcos2020}. See \url{www.changepoint.info} for an excellent review site and software repository on this topic. Said differently, change-points split the data into stationary subseries, see Fig. \ref{fig:window_selection}. This is what makes them interesting for model calibration and forecasting. 



Our algorithm for calibration window selection, i.e., identifying periods with similar time series dynamics to the currently observed, is as follows:
\begin{enumerate}
    \item Set the maximum number $N_c^{max}$ of change-points to be identified.
    
    \item Use the NOT method to identify $N_c \in [0, N_c^{max}]$ change-points $c_i,\ i = 1, \ldots, N_c$, in the initial calibration window $\vect{C}_0$ of length $\tau$. Additionally, denote the first observation in $\vect{C}_0$ by $c_0$.
    
    \item If $N_c=0$ return calibration sample $\vect{C}=\vect{C}_0$. Otherwise, compute the empirical quantiles $q_{low}$ and $q_{high}$ of the observations within the period between the most recent change-point found and the last observation in $\vect{C}_0$.
    
    \item For every interval between two subsequent change-points $c_{i-1}$ and $c_{i}$ compute the median $m_i$ of its observations, $i = 1, \ldots, N_c - 1$.
    
    \item Set $\vect{C} = \bigcup_{m_i \in (q_{low}, q_{high})}^i {[c_{i-1}, c_{i}]}$.
\end{enumerate}
Based on a limited simulation study, we set the maximum number of change-points $N_c^{max}=12$ and the order of quantiles $(q_{low}, q_{high})=(q_{0.025}, q_{0.975})$. We also use the least constraining form of NOT, i.e., we assume that the data have piecewise continuous variance and piecewise continuous mean. Any deviations from this are treated as a breach of stationarity.

\begin{figure}[tb]
\centering
\includegraphics[width=.9\textwidth]{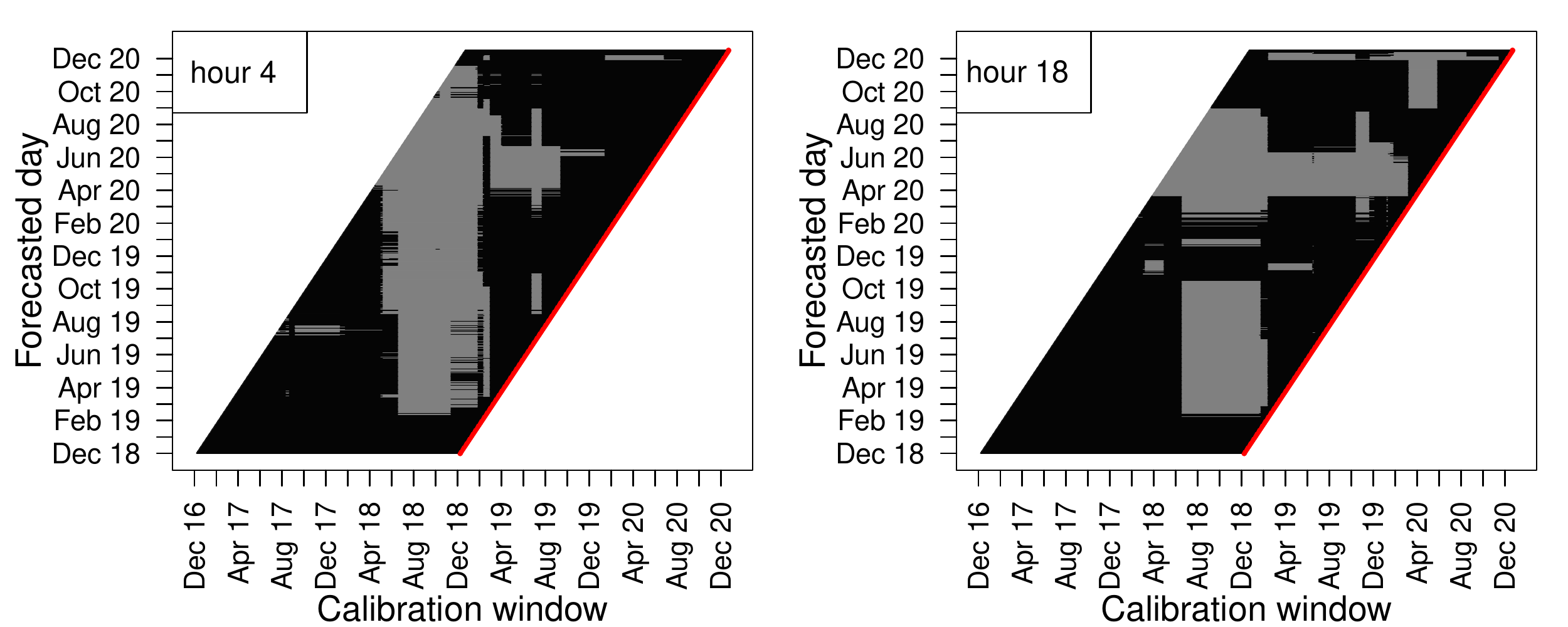}
\caption{Overview of NOT-selected (black) and discarded (gray) periods in the two-year calibration window (dates on the $x$ axis) for hours 4 (\textit{left panel}) and 18 (\textit{right panel}). The red line indicates the forecasted day (date on the $y$ axis).}
\label{fig:windows_de}
\end{figure}

A sample run of the algorithm is presented in Fig. \ref{fig:window_selection}. In the plot, the period closest to the forecasted day (red dot) is characterized by relatively stable, low prices with a small variance. The algorithm discards the light gray subperiods, when either the prices or their variance are significantly higher. 
In Figure \ref{fig:windows_de} we illustrate the results for two sample hours and the whole test window. We use a rolling scheme, i.e., once forecasts for the 24 hours of the first day in the test sample are computed, the calibration window is moved forward by one day and forecasts for the 2nd day in the test sample are calculated. A clear pattern of vertical gray stripes emerges, meaning that for a range of windows the change-points are consistently detected on the same or neighboring days. Comparing these plots with the price trajectory in Fig. \ref{fig:data_de}, we can observe that much fewer observations are selected by NOT when the prices tend to be more spiky, as can be seen in Spring 2020 (Apr 20 -- Jun 20 on the $y$ axis, esp. for hour 18).





\section{Forecasting models}\label{sec:methods}

For comparison purposes, the underlying model we use is the same as in \cite{nitka_forecasting_2021a}. It is an autoregressive structure with exogenous variables dubbed ARX. Since the prices $P_{d,h}$ are set in the day-ahead auction on day $d-1$ independently for the 24h of day $d$, it is customary in the EPF literature \cite{hubicka_note_2019a,lago_forecasting_2021} to treat every hour as a separate time series. Hence, we consider 24 ARX models of the form: 
\begin{equation}
\label{eq:ARX_P}
\begin{split}
P_{d,h}   = & \underbrace{\vect{\alpha}_h \vect{D}_{d}}_{\text{Dummies}}+\underbrace{\sum_{p \in \{1,2,7\}} \beta_{h,p} P_{d-p,h}}_{\text{AR component}}+\underbrace{\theta_{h,1}P_{d-1,min} +\theta_{h,2}P_{d-1,max}}_{\text{Yesterday's price range}}\\
& +\underbrace{\theta_{h,3}P_{d-1,24}}_{\text{Last known price}} +\underbrace{\theta_{h, 4}\hat{L}_{d,h} + \theta_{h, 5}\hat{W}_{d,h} + \theta_{h, 6}\hat{S}_{d,h}}_{\text{Exogenous variables}} +\ \! \underbrace{\varepsilon_{d,h}}_{\text{Noise}}.
\end{split}
\end{equation}
The autoregressive (AR) dynamics are captured by the lagged prices from the same hour yesterday, two and seven days ago. Following \cite{ziel_dayahead_2018a}, yesterday's minimum $P_{d-1,min}$, maximum $P_{d-1,max}$ and the last known price $P_{d-1,24}$, as well as day-ahead predictions of the three exogenous variables are included. Finally, a $1\times7$ vector of dummy variables $\vect{D}_{d}$ is used to represent the weekly seasonality and the uncertainty is represented by white noise.



Overall, we compare seven types of approaches that all use ARX as the underlying model. The first three are the same as in \cite{nitka_forecasting_2021a}: 
(i) \textbf{Win($\tau$)} -- the ARX model estimated using a window of $\tau$ days, with $\tau \in [56,57,...,728]$, 
(ii) \textbf{Av(Win)} -- the arithmetic average of six forecasts of the ARX model for three short ($\tau=56, 84, 112$) and three long windows ($\tau=714, 721, 728$), and 
(iii) the \textbf{ARHNN} model. The next four include: 
(iv) \textbf{Win\textsubscript{H}($\tau$)} -- the same as Win($\tau$) but calibrated to \textit{asinh}-transformed prices, 
(v) \textbf{NOT\textsubscript{H}(728)} -- the ARX model calibrated to \textit{asinh}-transformed prices in NOT-selected subperiods from the 728-day window, 
(vi) \textbf{Av(Win\textsubscript{H})} -- the same as Av(Win) but calibrated to \textit{asinh}-transformed prices, and 
(vii) \textbf{Av(NOT\textsubscript{H})} -- the same as Av(Win\textsubscript{H}) but with the forecasts for the three long windows ($\tau=714, 721, 728$) replaced by NOT\textsubscript{H}(728). The rationale behind the latter averaging scheme is that NOT\textsubscript{H}($\tau$) performs best for long calibration windows and offers little or even no gain for $\tau <1$ year.

\section{Results}\label{sec:results}

\begin{figure}[tb]
    \begin{minipage}{0.4\textwidth} 
        \centering
        \begin{tabularx}{0.8\textwidth}{X|X}
        \hline
            \textbf{Method} & \textbf{RMSE} \\ \hline \hline
            Win(728) & 8.2860 \\
            Av(Win) & 8.0286 \\
            ARHNN & 7.8605 \\
            Win\textsubscript{H}(728) & 7.7286 \\
            NOT\textsubscript{H}(728) & 7.5994 \\
            Av(Win\textsubscript{H}) & 7.0968 \\
            Av(NOT\textsubscript{H}) & 7.0831 \\
            \hline
        \end{tabularx}
    \end{minipage} 
    \begin{minipage}{0.55\textwidth} 
        \centering  
        \includegraphics[width=0.8\textwidth]{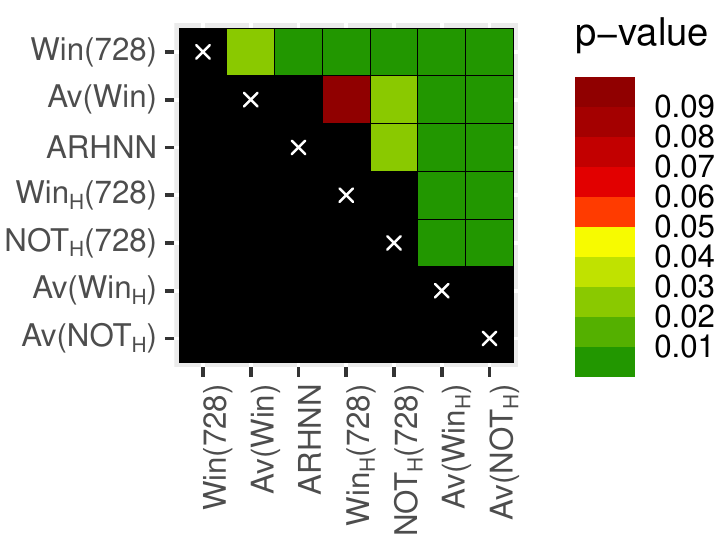}
    \end{minipage}
    \caption{RMSE errors in the out-of-sample test period (\textit{left panel}). A heatmap of the $p$-values for the multivariate Diebold-Mariano test \cite{ziel_dayahead_2018a} for each pair of methods (\textit{right panel}). The smaller the $p$-values, the more significant is the difference between the forecasts of a model on the $x$-axis (better) and the forecasts of a model on the $y$-axis (worse). Black color indicates $p$-values in excess of 0.1.
    }\label{fig:heatmap}
\end{figure}

We evaluate the forecasting performance of the seven approaches presented in Section \ref{sec:methods} in terms of the \textit{root mean squared error} (RMSE; results for the mean absolute error are similar and available from the authors upon request). The RMSE values reported in Fig.\ \ref{fig:heatmap} are aggregated (averaged) across all hours in the 736-day test sample, see Fig.\ \ref{fig:data_de}. Additionally, to test the significance of differences in forecasting accuracy, for each pair of models we employ the multivariate variant of the Diebold-Mariano (DM) test, as proposed in \cite{ziel_dayahead_2018a}.

Several conclusions can be drawn. Firstly, changing the preprocessing method from normalization \cite{nitka_forecasting_2021a} to \textit{asinh} transformation \cite{uniejewski_variance_2018} generally reduces the RMSE. 
Even the worst performing out of the latter approaches, Win\textsubscript{H}(728), improves on ARHNN, the most accurate method in \cite{nitka_forecasting_2021a}. Secondly, NOT-selection yields further improvement, although not statistically significant if considered on its own. Compare NOT\textsubscript{H}(728) with Win\textsubscript{H}(728) and Av(NOT\textsubscript{H}) with Av(Win\textsubscript{H}).

\begin{figure}[tb]
\includegraphics[width=\textwidth]{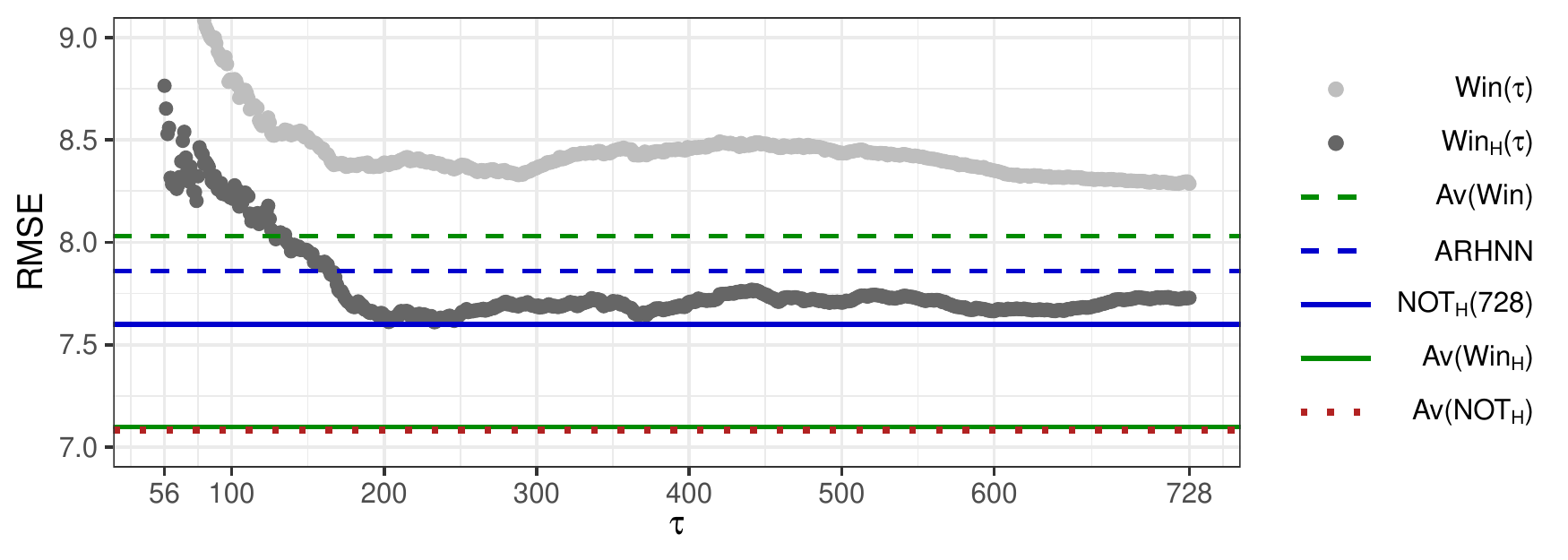}
\caption{RMSE values for all considered models; $\tau$ is the calibration window length.} 
\label{fig:rmse}
\end{figure}

The RMSE values for all considered approaches are presented graphically in Figure \ref{fig:rmse}. It clearly shows the significant improvement from using the \textit{asinh} -- compare between Win($\tau$) with Win\textsubscript{H}($\tau$) for all $\tau$'s. While Win\textsubscript{H}(728) is not the best performing of all Win\textsubscript{H}($\tau$) models, the differences in performance are relatively minor for $\tau \geq 200$ days. Even the best ex-post known model, Win\textsubscript{H}(233), is slightly worse than NOT\textsubscript{H}(728). The averaged forecasts Av(Win\textsubscript{H}) and Av(NOT\textsubscript{H}) are further able to improve on the accuracy, although the differences between them are not significant.



\section{Conclusions and discussion}\label{sec:conclusions}

In this paper we propose a novel method for selecting calibration subperiods based on Narrowest-Over-Threshold (NOT) change-point detection \cite{baranowski_NOT_2019}. Contrarily to the traditional time series approach in which the most recent observations are taken as the calibration sample, we propose to estimate the predictive models only using data in the selected subperiods.
We evaluate our approach using German electricity market data and seven variants of autoregressive models tailored for electricity price forecasting (EPF). We provide empirical evidence that significant improvement in forecasting accuracy can be achieved compared to commonly used EPF approaches, including the recently proposed ARHNN \cite{nitka_forecasting_2021a}.
In addition to calibration sample selection, our results also emphasize the importance of using transformations like the \textit{asinh}, in line with \cite{uniejewski_variance_2018,ziel_dayahead_2018a}. 

The roughly sixfold increase in computational time of the NOT-based methods -- 4.97s for 24 hourly forecasts using NOT\textsubscript{H}(728)  vs. 0.83s using Win\textsubscript{H}(728), running R ver.\ 3.6.3 on an i7-9750H processor --  can be seen as a drawback, especially compared to less complex ways of improving forecast accuracy, like calibration window averaging \cite{hubicka_note_2019a}. 
However, the automation of the forecasting process may make the trade-off worthwhile. If this is the case for more complex models than the autoregressive ones considered here or the shallow neural network in \cite{DeMarcos2020}, e.g., LASSO-estimated AR (LEAR) and deep neural networks \cite{lago_forecasting_2021}, is left for future work.


%
%
%
\bibliographystyle{splncs04}
\bibliography{NasiadkaNitkaWeron22_ICCS}
\end{document}